\documentclass[11pt,twoside]{article}

\usepackage{asp2004}
\usepackage{epsf}
\usepackage{lscape}
\usepackage{graphicx}

\newcommand{\etal}{et al.}
\newcommand{\teff}{{\mathrm{T}}_{\mathrm{eff}}}
\newcommand{\zzc}{ZZ~Ceti}

\begin{document}
\title{Mean \zzc\ pulsation period gauges stellar temperature}   
\author{Anjum S. Mukadam\altaffilmark{1}}   
\affil{Department of Astronomy, University of Washington, Seattle, WA\,-\,98195-1580, USA}    
\altaffiltext{1}{Hubble fellow}
\author{M. H. Montgomery, A. Kim, D. E. Winget}
\affil{Department of Astronomy, University of Texas at Austin, Austin, TX\,-\,78712, USA}
\author{S. O. Kepler}
\affil{Instituto de F\'{\i}sica, Universidade Federal do Rio Grande do Sul, 91501\,-\,970 Porto Alegre, RS\,-\,Brazil}
\author{\& J. C. Clemens}
\affil{Department of Physics and Astronomy, University of North Carolina, Chapel Hill, NC 27599, USA}

\begin{abstract} 
The mean pulsation period of \zzc\ stars increases with decreasing effective temperature
as we traverse from the blue to the red edge of the instability
strip. This well-established correlation between the mean period
and spectroscopic temperature suggests that the mean period could be utilized as a
tool to measure the relative temperature of the star independent of spectroscopy.
Measuring the pulsation periods of a \zzc\ star is a simple,
model-independent, and straight forward process as opposed to a
spectroscopic determination of its temperature.

Internal uncertainties in determining the spectroscopic temperature of a \zzc\
star are at least 200\,K, 15\% of the 1350\,K width of the instability strip.
The uncertainties in determining the mean period arise mostly from amplitude modulation
in the pulsation spectrum and are smaller than 100\,s for 91\% of the \zzc\ stars,
$<$8\% of the 1300\,s width of the
instability strip. In principle this implies that for 90\% of the \zzc\ stars, the average uncertainty
in determining the location of a \zzc\ star within the instability
strip decreases by a factor of two in utilizing
the mean period of the \zzc\ star as a temperature indicator rather than conventional
spectroscopy. Presently we only claim that
the relative temperatures of \zzc\ stars derived by using the mean pulsation period are
certainly as good as and perhaps about 15\% better than spectroscopy.
\end{abstract}

\section{Introduction}
Non-interacting hydrogen atmosphere (DA) white dwarfs
pulsate in a narrow instability strip located within the temperature range
10800--12300\,K for $\log~g\approx~8$ \citep{Bergeronet95,Bergeronet04,
KoesteraAllard00,Mukadamet04,Gianninaset05}.
These DA variables (DAVs) are also called the \zzc\ stars.
The hot \zzc\ stars show
relatively few pulsation modes, with low amplitudes ($\sim $0.1--3\%) and
periods around 100--300\,s \citep[see][]{Clemens93}.
The cooler DAVs show longer periods, around 600--1000\,s, larger amplitudes (up to 30\%),
non-linear pulse shapes,
and greater amplitude modulation \citep[e.g.][]{Kleinmanet98}.
The pulsation characteristics of the hot and cool DAVs are quite distinct, and have
allowed us to classify these stars meaningfully.

\section{Weighted Mean Period as a function of effective temperature}
\citet{Clemens93} was the first to systematically demonstrate the increase
in mean pulsation period as a function of decreasing temperature for a significant
sample of the \zzc\ stars. With the discovery of additional \zzc\ stars, this empirical
correlation was confirmed again by \citet{Kanaanet02}
and more recently by \citet{Mukadamet06}.

Using only the independent excited modes and excluding
any harmonics and linear combinations, we compute the mean period for a given pulsation spectrum,
weighting each period with the corresponding amplitude. In case of multiple pulsation spectra for
a given \zzc\ star, we determine an average of the individual values.
We show the weighted mean period (WMP) as a function
of spectroscopic temperature for two independent samples of \zzc\ stars, and find that the
weighted mean pulsation period correlates
well with the spectroscopic temperature in both cases. Note that we do not claim that the relationship between the WMP
and the temperature is linear; a straight line is
merely the simplest model fit possible to the
observations shown in Figure 1, considering the observed amount of scatter.
 
\begin{figure}[!ht]
\centering
\includegraphics[height=2.5in,clip=true]{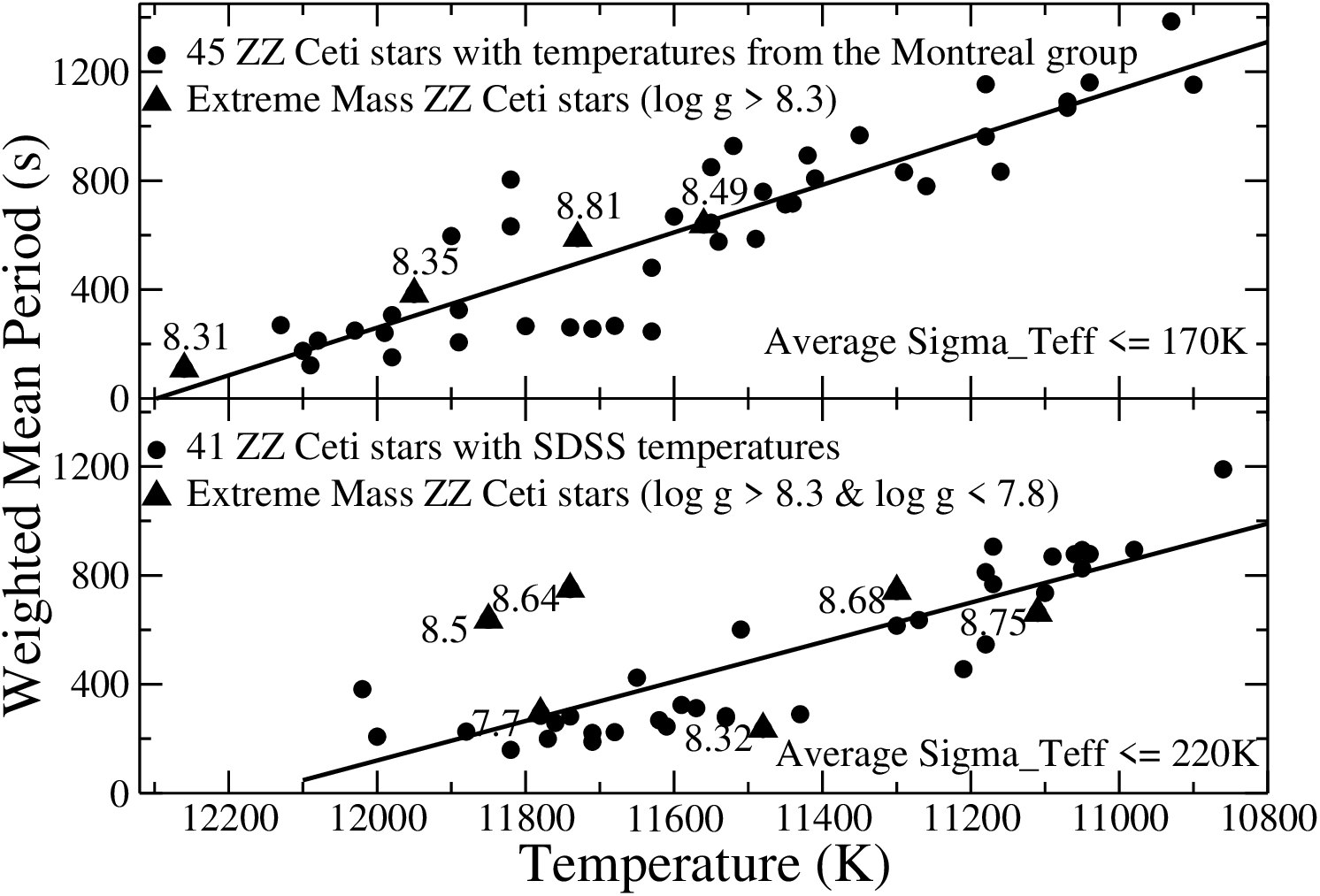}
\caption{The weighted mean period (WMP) correlates well with spectroscopic temperature for two independent samples
of \zzc\ stars. The top panel shows
45 DAVs with temperatures from \citet{Bergeronet04}, and \citet{Gianninaset05,Gianninaset06}.
The bottom panel shows 41 SDSS DAVs with temperatures from the DR4 DA white dwarf catalog.}
\end{figure}

The scatter, as measured along the x-axis, helps us constrain the average uncertainty in
spectroscopic temperature, provided we assume that most of the scatter in the correlation arises due to
uncertainties in temperature. We hereby constrain the average uncertainty in temperature
for the 45 \zzc\ sample published by \citet{Bergeronet04} and \citet{Gianninaset05,Gianninaset06}
to be $\leq$170\,K. We also constrain the average uncertainty in temperature for the 41 SDSS \zzc\
sample to be $\leq$220\,K, which we acquired from the DR4 catalog for DA white dwarfs\footnote{
http://www.naoj.org/staff/sjnk/varcat/wdDAVARS.tableRA.DR4.html}, 
based on version auto22 of the spectral fitting algorithm.

The 45 \zzc\ star sample shown in the top panel is our best case scenario; the temperatures
come from high S/N optical spectra and most of these bright stars have well determined pulsation spectra,
have been observed multiple times, and sometimes even from multiple sites.
For this sample, we find that our upper limit of 170\,K for the average uncertainty in
spectroscopic temperatures is {\bf lower} than the estimate of 200\,K from
\citet{Bergeronet04}. This is suggestive that the correlation between WMP and
spectroscopic temperature is tighter than it seems, and that most of the scatter arises due to
the uncertainties in temperatures, that have been derived from optical spectra.

Lastly, we highlight the extreme mass stars with $\log~g>8.3$ or $\log~g<7.8$
in Figure 1 as upright triangles. For our best-case
bright 45 \zzc\ sample, we find that the massive stars are not outliers in the
WMP-$\teff$ correlation. The massive crystallized pulsator BPM\,37093 with $\log~g=8.81$, which
has been the subject of a WET run, agrees fairly well with the linear fit
\citep{Kanaanet05}.
Three of the six extreme mass stars from the SDSS DAV sample
are outliers in the plot, but most of these stars have pulsation spectra determined
from no more than a few hours of discovery data. Also, the masses of the SDSS \zzc\ stars
have a relatively higher uncertainty than the Montreal sample as they have been determined
from lower S/N optical spectra.
However the behaviour of the bright massive stars, and specially BPM\,37093,
is surprising. We expect that at a given temperature, white dwarfs of different masses have
convection zones of different thicknesses. Our models suggest that the driving frequency is governed
by the thermal timescale at the base of the convection zone. Massive white dwarfs
should show a mean pulsation period longer than other average mass
DAVs at the same temperature. The empirical WMP values for all four massive stars in the Montreal group sample,
and four out of six extreme mass \zzc\ stars in the SDSS sample contradict this theoretical expectation.
Although we presently have a small number of extreme mass \zzc\ stars, their WMP values have
implications for the slope of the blue edge and pulsation models in general.

The observed correlation between WMP and $\teff$ suggests that WMP can be used to gauge stellar
temperature. Determining the pulsation periods of a \zzc\ star involves small uncertainties,
and is a relatively simple model-independent process.
We cannot determine an absolute temperature using WMP; we can only derive the position of a
\zzc\ star within the instability strip. To do so, we do not need the correlation shown in
Figure 1. As long as we restrict our relative $\teff$ parameter
in units of seconds in the WMP temperature scale, the uncertainty in
our measurements is not related to spectroscopic temperatures.
It is only when we attempt to translate WMP into a temperature in degrees, that
we have to use the relation between WMP and $\teff$ from Figure 1. Even in this case,
we are better off than the typical 200\,K uncertainty in spectroscopic temperatures
because the slopes of the best fit lines in Figure 1 depend
on 40--45 stars.
Using WMP directly as a temperature scale is non-intuitive at the present time.
But this maybe worth thinking about as an alternative scale in the long run, as we
improve our understanding of the relation between WMP and $\teff$, both observationally and theoretically.

\section{Exploring the range of excited periods vs. temperature}
We show the longest (middle panel) and shortest (bottom panel)
excited periods for our best-case bright 45 \zzc\ sample as
a function of spectroscopic temperatures from the Montreal group in Figure 2. We also
include a plot of WMP vs. temperature (top panel) for comparison. We indicate a measure
of the scatter in each of the panels, and find that the shortest excited period exhibits
the tightest correlation with $\teff$. But this may or may not be significant; the SDSS
sample, for which we have no more than a few hours of discovery data, does not mimic
this general behaviour.

\begin{figure}[!ht]
\centering
\includegraphics[height=2.5in,clip=true]{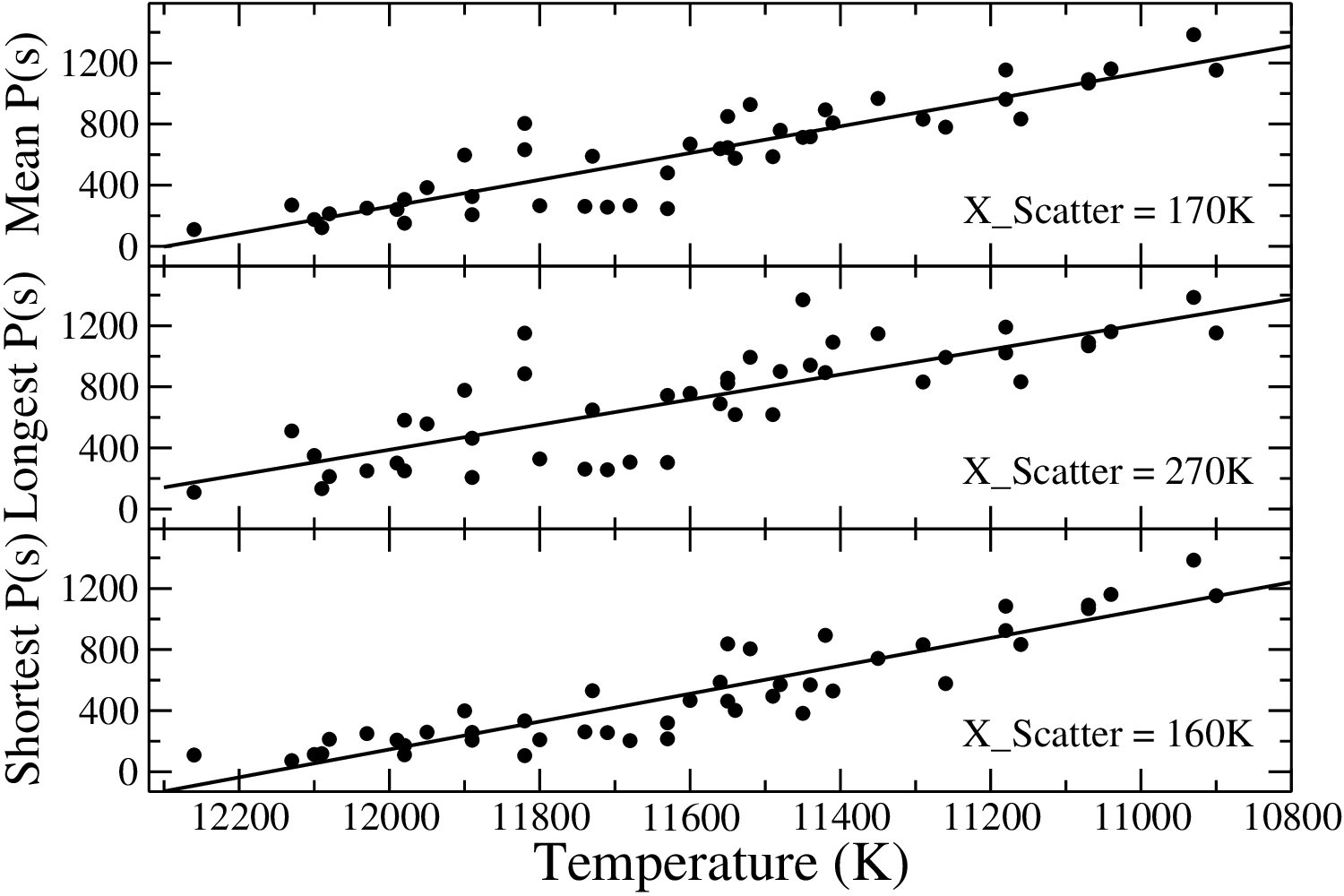}
\caption{Shortest and longest excited periods as a function of spectroscopic temperature,
plotted for comparison with the weighted mean period (WMP).}
\end{figure}

	We determine WMP using all the independent periods excited in the star, but the weighting
is amplitude dependent, which in turn depends on the inclination angle.
Using the shortest or the longest period, we overcome the amplitude dependence of WMP, but find ourselves
relying on a single period determination instead. 
Some of the older pulsation spectra come from data acquired
using one and two channel photometers, which could not be used reliably to separate 
long pulsation periods from sky variations, unless the data came from South Africa.
Also, the determination of the
shortest pulsation period depends on the time resolution of the data. Since we have a highly
diverse set of pulsation spectra with stars of different magnitudes and pulsation amplitudes
observed using different telescopes, we hesitate to draw any implications
from Figure 2.

\section{Amplitude Modulation: Dominant source of uncertainty in WMP}
The cool \zzc\ stars exhibit a significant amount of amplitude modulation, and their pulsation
spectra can change substantially from one season to the next \citep[e.g.][]{Kleinmanet98}.
We believe this is the largest source of uncertainty in determining WMP and in using it as a temperature scale. 
In order to estimate the size of this source of uncertainty, we select those \zzc\ stars
in our data set that have multiple seasons of observations. We then determine the spread in the WMP values
computed from the multiple pulsation spectra. For 36 \zzc\ stars, we show the 1\,$\sigma$ departure of WMP from
the average obtained using multiple seasons of observations in Figure 3. We also show
the worst case scenario, the maximum deviation of WMP from the average value.
We find that 92\% of the sample show a 1\,$\sigma$ deviation smaller than 50\,s, and
the whole sample shows a modulation in WMP smaller than 80\,s.
Even in the worst case, about 91\% of the 36 \zzc\ stars exhibit a
modulation in WMP smaller than 100\,s, and 97\% show a modulation smaller than 130\,s.

\begin{figure}[!ht]
\centering
\includegraphics[height=2.5in,clip=true]{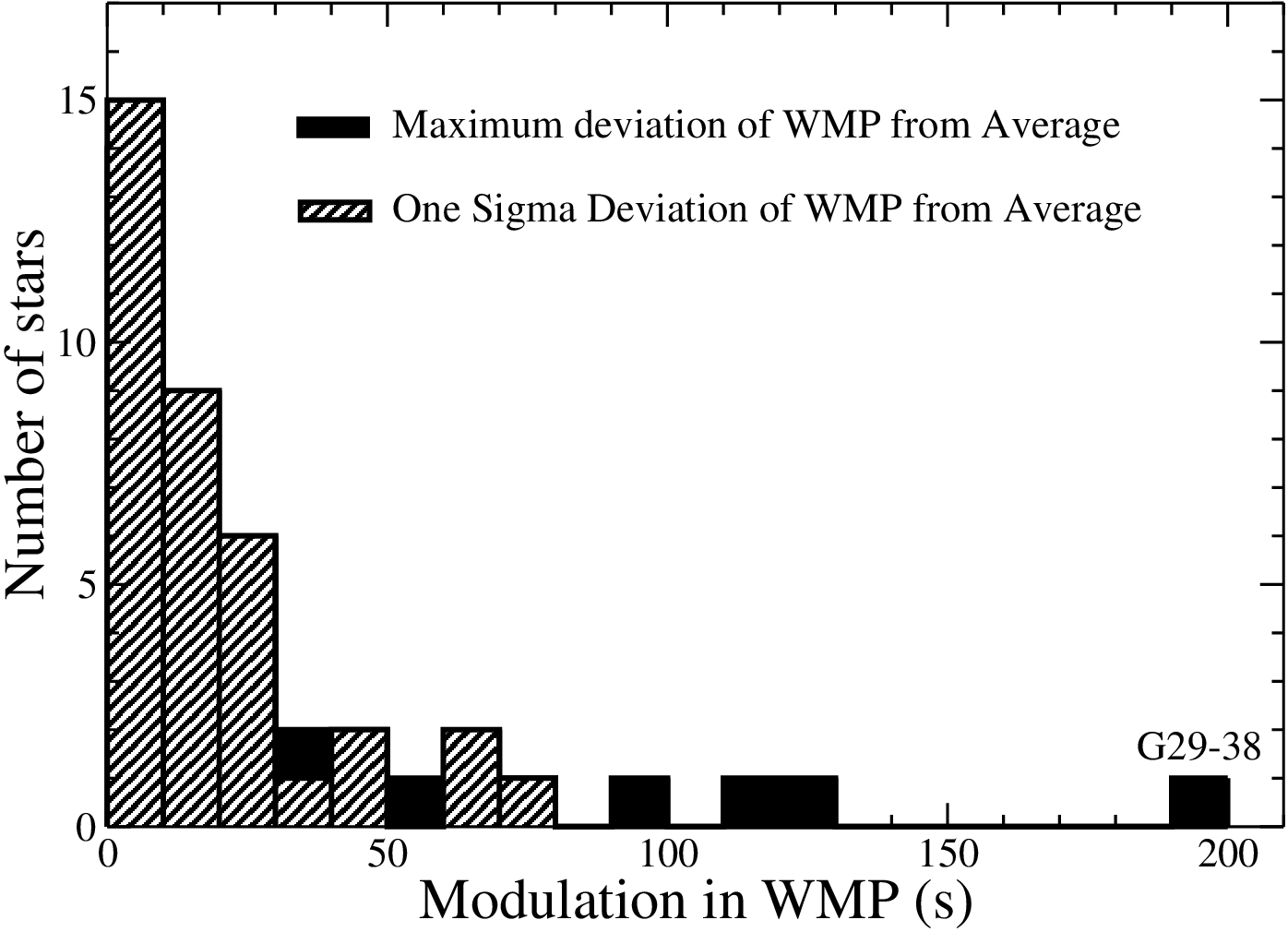}
\caption{Using a sample of 36 \zzc\ stars, we demonstrate that amplitude modulation results
in a change in weighted mean period (WMP) smaller than 100\,s even in the worst case scenario
for about 91\% of the sample.}
\end{figure}

Adopting the worst-case numbers, we find that the uncertainty in WMP is likely to be
smaller than 100\,s for about 91\% of the \zzc\ stars, which translates to $<$8\% of the 1300\, s width of
the instability strip in period. Adopting the best-case scenario for conventional spectroscopy,
we will use our limit of 170--220\,K as the average uncertainty in spectroscopic
temperatures to arrive at an uncertainty of 13--16\% of the 1350\,K width of the instability strip.
This suggests that for 90\% of the \zzc\ stars, we can improve our knowledge of the
star's location
within the instability strip by a factor of two, by using WMP instead of spectroscopy. 
However we have to face some theoretical questions before we can be certain of the factor
of improvement in using WMP instead of spectroscopy.

\section{Chicken or Egg? Mean Period or Spectroscopic Temperature?}
To determine which temperature scale is better, WMP or spectroscopic $\teff$, we have
devised the following test.
The observed mean pulsation amplitude of \zzc\ stars shows a large scatter when plotted as a function of
$\teff$ \citep[e.g.][]{Kanaanet02,Mukadamet06}, and this is largely due to the different inclination angles.
If all DAVs with the same $\teff$ had the same inclination angle, then we theoretically expect to see
a tight correlation between the mean amplitude and temperature. This is not strictly true because
pulsation amplitude also depends on the mass of the star; massive \zzc\ stars have small amplitudes.
Excluding the massive DAVs ($\log~g\geq8.5$) and the low mass DAVs ($\log~g\leq7.7$) from both
samples, we compute the scatter in plotting mean amplitude as
a function of $\teff$ and also as a function of WMP. The temperature scale that was worse than the
other would introduce an additional scatter in the plot. Having conducted the test, we find a 15\%
reduction in scatter when using WMP as a temperature scale instead of spectroscopic $\teff$ for both samples.
This suggests that WMP is at least 15\% better than traditional spectroscopy.

\section{A question to ponder}
Montgomery (2005)
show that the thermal timescale at the base of the convection zone $\tau_{th}$ is proportional
to $\teff^{-90}$ and have successfully fit the non-linear pulse shape of one DAV to date, namely G29-38.
\citet{Kimet06} find that the empirical WMP is not consistent with
the thermal timescale at the base of the convection zone in models.
The question to ask is: what physical quantity in the star
does WMP represent. Unless we can comprehend this satisfactorily, we will
not be able to understand the tightness of its correlation with $\teff$ and the
true uncertainties involved in using this technique to determine
stellar temperature.

\acknowledgements 
Support for this work was provided by NASA through the Hubble Fellowship grant
HST-HF-01175.01-A awarded by the Space Telescope Science Institute, which is operated
by the Association of Universities for Research in Astronomy, Inc., for NASA, under contract
NAS 5-26555.


\end{document}